\documentstyle[11pt]{article}
\newcommand{\be}{\begin{equation}}
\newcommand{\en}{\end{equation}}
\newcommand{\bea}{\begin{eqnarray}}
\newcommand{\ena}{\end{eqnarray}}
\newcommand{\dps}{\displaystyle}
\newcommand{\dcon}[1]{\raisebox{-9.5mm}[1ex][-9.5mm]
                      {$\stackrel{\displaystyle #1}{\unoc}$}}
\newcommand{\ducon}[1]{\raisebox{-9.5mm}[1ex][-9.5mm]
                      {$\stackrel{\displaystyle #1}{\uunoc}$}}
\newcommand{\scon}[1]{\raisebox{-7.5mm}[1ex][-7.5mm]
                      {$\stackrel{\displaystyle #1}{\snoc}$}}
\newcommand{\snoc}{\rule[5mm]{0.2mm}{1mm}\rule[5mm]{1cm}{0.2mm}
                  \rule[5mm]{0.2mm}{1mm}}
\newcommand{\unoc}{\rule[5mm]{0.2mm}{3mm}\rule[5mm]{1.3cm}{0.2mm}
                  \rule[5mm]{0.2mm}{1.5mm}}
\newcommand{\uunoc}{\rule[5mm]{0.2mm}{1.5mm}\rule[5mm]{1.3cm}{0.2mm}
                     \rule[5mm]{0.2mm}{3mm}}

\newcommand{\vs}[1]{\rule[ - #1 mm]{0 mm}{#1 mm}}
\newcommand{\sm}[2]{\frac{\mbox{\footnotesize #1}\vs{-2}}
                   {\vs{-2}\mbox{\footnotesize #2}}}
\newcommand{\SC}{{\cal C}}
\newcommand{\W}{{\sf W}}
\newcommand{\NP}[1]{Nucl.\ Phys.\ {\bf #1}}
\newcommand{\PL}[1]{Phys.\ Lett.\ {\bf #1}}

\newcommand{\MPL}[1]{Mod.\ Phys.\ Lett.\ {\bf #1}}
\newcommand{\IJMP}[1]{Int.\ J.\ Mod.\ Phys.\ {\bf #1}}

\newcommand{\sect}[1]{\setcounter{equation}{0}\section{#1}}

\begin{document}
%
%
\renewcommand{\thefootnote}{\fnsymbol{footnote}}
\newpage
\setcounter{page}{0}
\pagestyle{empty}
%
%
\rightline{KCL-TH-92-9}
\rightline{DFTT-70/92}
\rightline{hep-th/9212104}
\rightline{December 1992}

\vs{15}

\begin{center}
{
\LARGE {\W-algebras with set of primary fields of dimensions $(3, 4, 5)$ and
$(3,4,
5,6)$ }\footnote{Work in parts supported by the
UK Science and Engineering Research Council.
} }\\[1cm]
{\large K.\ Hornfeck}\footnote{e-mail: HORNFECK@TO.INFN.IT; \hspace{0.5cm}
31890::HORNFECK}\\[0.5cm]
{\em King's
College, Department of Mathematics, Strand,
London WC2R 2LS, GB}\\
{\em INFN, Sezione di
Torino, Via Pietro Giuria 1, Italy} \footnote{Address after October 1st, 1992.}
\\[1cm]
\end{center}
\vs{15}

\centerline{ \bf{Abstract}}
We show that that the  Jacobi-identities for a \W-algebra with primary fields
of
dimensions 3, 4 and 5  allow two different  solutions. The first solution can
be
identified  with {\sf  WA}$_4$. The  second is  special in the  sense that,
even
though associative for  general value of the central  charge, null-fields
appear
that violate  some of the  Jacobi-identities, a  fact that is  usually linked
to
exceptional \W-algebras.
In contrast we find for the algebra that has an additional spin 6 field only
the
solution {\sf WA}$_5$.
%
%
\renewcommand{\thefootnote}{\arabic{footnote}}
\setcounter{footnote}{0}
\newpage
\pagestyle{plain}
\sect{Introduction}
Over the last couple of years, a varity of different \W-algebras has been
investigated~\footnote{For a recent review on \W-algebras and extensive list
of references see~\cite{BS92}.}.

The efforts of classifying \W-algebras~\cite{FRS92,FOR91,BW91}
led to a better
understanding of the origin of \W-algebras and to an
increasing number of \W-algebras.
However, some \W-algebras still lack an explanation. We shall present here
another example, appearing when we consider the possible \W-algebras with a
set of primary fields of dimension 3, 4 and 5. In the following we denote a
\W-algebra with a set of primary fields of dimensions $d_1, d_2, \ldots$ {\it
in
addition} to the stress energy tensor $T$ as
\W($2, d_1, d_2, \ldots$).

We want to  construct these
\W-algebras  by demanding  that the Jacobi-identities are  satisfied.
We therefore check that the
Jacobi-identities that are given for any three simple fields $A$, $B$ and $C$
as the function of double-contractions
\be
\mbox{{\sc Jacobi}}[A,B,C] :=
\dcon{A(x)\,\scon{B(y)\,C(z)}}
- \epsilon_{AB} \dcon{B(y)\,\scon{A(x)\,C(z)}}
- \ducon{\scon{A(x)\,B(y)}\,C(z)} \;-\;O((x-z)^o)
\;-\;O((y-z)^o)
\label{jac}
\en
``vanish''.  Here  $\epsilon_{AB} =  -1$ if $A$  and $B$  anticommute and
$+1$ else and  only poles  in $(y-z)$  and $(x-z)$ are considered and the poles
in $(x-y)$ that appear in the last contraction of the definition~(\ref{jac})
are
expanded in poles in $(x-z)$ and zeros in $(y-z)$,
\be
\frac{1}{x-y} \,=\, \frac{1}{x-z}\, \sum_{l=0}\left(\frac{y-z}{x-z}
\right)^l
\en

One has to be  cautious with the meaning  of the word  ``vanish''. As
Zamolodchikov  has shown in his  original paper~\cite{Zam86}, there are
\W-algebras, where the  Jacobi-identities are violated by the apearance of
null-fields. The simplest example for this is the algebra built by a  primary
fermionic field of  dimension 5/2, $V$, and  the stress energy tensor
$T$~\cite{Zam86}. It turns out that only for $c=-13/14$ the simple field $V$
itself disappears  from the Jacobi-identity $\mbox{{\sc Jacobi}}[V,V,V]$ and
even
then
the descendent of $V$ at level $3$,
\be
-\sm{4}{9}\,:T \partial V: + \sm{5}{9}\,:\partial T V: +
\sm{4}{63}\,\partial^3 V
\en
still appears.  However, this field  is a null-field  exactly for the  same
central charge  $c=-13/14$. This  behaviour is a common characteristic for the
exceptional \W-algebras, i.e.\ those \W-algebras that are only defined
for a finite set of central charges (though for some of the allowed values of
central charges for an exceptional \W-algebra no null-field might be involved).

Therefore we should use ``vanish'' in the sense that that the Jacobi-identity
$\mbox{{\sc Jacobi}}[A,B,C]$ is zero modulo null-fields.

The choice (\ref{jac}) for the Jacobi-identities has the advantage that they
are
expressed
in terms of fields rather than their modes, so unnecessary combinatoric
can be avoided.
In our
computations we  use the {\it Mathematica}$^{\mbox{\tiny TM}}$ package for
computing OPEs by  K.~Thielemans~\cite{Thi91,Mat91} in which
the function $\mbox{{\sc Jacobi}}[A,B,C]$ can be easily defined.

\sect{Solutions to the algebra \W$(2,3,4,5)$}
There are ten non-trivial Jacobi-identities to satisfy:
{\sc Jacobi}[$W_n$, $W_m$, $W_l$] for $n \geq m \geq l$.
The Jacobi-identities $\mbox{{\sc Jacobi}}$[$T$, $T$,
$W_n$] are automatically satisfied when the fields
$W_n$ are primary and {\sc Jacobi}[$T$, $W_n$, $W_m$]
when the ansatz for the OPE  $W_n(z) \, W_m(w)$ has already the right conformal
structure, i.e.\
primary fields $P$
appear together
with their descendents with the right conformal factors. Unknown parameters
are only the
structure constants $\SC_{n,m}^P$. As
primary  fields will however not only appear the simple fields, but also
primary
fields that are constructed out of (normal ordered) products of the simple
fields. For these composite fields we shall use the notation
\bea
P^{n,m}_d &= &\frac{1}{(d-n-m)!} :W_n^{(d-n-m)}\, W_m :\,+ \;
              \mbox{descendents of primary fields (simple}\label{prf}
\nonumber \\
&&\hspace{5.8cm} \mbox{or not) of lower dimension}
\ena
In the algebra we are considering here, primary composite
fields up to dimension 8
can appear  in the OPE of the simple fields. So we take the ansatz (using
standard-normalisation and omitting already those terms that will not
contribute
by symmetry arguments)
\bea
W_3 \star W_3 & = & \sm{c}{3}\, I + \SC^4_{3,3}\,  W_4 \label{ans33}\\[2mm]
W_3 \star W_4 & = & \SC^3_{3,4} \, W_3 + \SC^5_{3,4}\,  W_5
\label{ans34}\\[2mm]
W_3 \star W_5 & = & \SC^4_{3,5} \, W_4 + \SC^{P^{3,3}_6}_{3,5} \, P^{3,3}_6
\label{ans35}\\[2mm]
W_4 \star W_4 & = & \sm{c}{4} \, I + \SC^4_{4,4} \, W_4 +
                     \SC^{P^{3,3}_6}_{4,4} \, P^{3,3}_6 \label{ans44}\\[2mm]
W_4 \star W_5 & = & \SC^3_{4,5} \, W_3 + \SC^5_{4,5} \, W_5 +
                     \SC^{P^{3,4}_7}_{4,5} \, P^{3,4}_7 +
                     \SC^{P^{3,4}_8}_{4,5} \, P^{3,4}_8\label{ans45}\\[2mm]
W_5 \star W_5 & = & \sm{c}{5} \, I + \SC^4_{5,5} \, W_4 +
                     \SC^{P^{3,3}_6}_{5,5} \, P^{3,3}_6 +
                     \SC^{P^{3,3}_8}_{5,5}\,  P^{3,3}_8 +
                     \SC^{P^{3,5}_8}_{5,5} \, P^{3,5}_8 +
                     \SC^{P^{4,4}_8}_{5,5} \, P^{4,4}_8
\label{ans55}
\ena
The  Jacobi-identities  now impose  stringent  constraints on  the structure
constants $\cal C$.  Indeed, only
two sets of  non-equivalent solutions finally obey all
Jacobi-identities.

For the set of simple fields we are considering, the constants $\SC^l_{mn}$ and
$\SC^n_{ml}$ are simply related by~\cite{BFK91}
\be
\SC^l_{mn} \, = \, \sm{$l$}{$n$} \, \SC^n_{ml}
\en
and for both solutions we have therefore
\be
 \SC^3_{3,4}  =  \sm{3}{4} \SC^4_{3,3} \hspace{1cm}
 \SC^4_{3,5}  =  \sm{3}{5} \SC^5_{3,4} \hspace{1cm}
 \SC^3_{4,5}  =  \sm{3}{5} \SC^5_{3,4} \hspace{1cm}
 \SC^4_{5,5}  =  \sm{4}{5} \SC^5_{4,5} \hspace{1cm}
\en

The other structure constants for the first solution to the Jacobi-identities
are listed in Table I.
The squares in the definitions of $\SC^4_{3,3}$ and  $\SC^5_{3,4}$ and the
dependency of some of the constants on $\SC^4_{3,3}$ and $\SC^5_{3,4}$ reflect
the symmetries $W_n \rightarrow - W_n$.

The existence of this  solutions was already  known, though the form of the
structure constants has  not yet been given: It can be identified with the {\sf
WA}$_4$-algebra that has exactly the  set of simple fields we consider. This
identification can be made by the structure constants $\SC^4_{3,3}$ and
$\SC^4_{4,4}$:  In~\cite{Hor92}, these constants have been derived for {\sf
WA}$_{n-1}$ for any $n > 3$, using the free field  realisation of Fateev and
Lykyanov for these  algebras~\cite{FL88}~\footnote{These constants are for $n
\rightarrow \infty$ identical of those of the \W$_\infty$-algebra of
Pope~\cite{PRS90}, after changing their basis of simple fields to the one used
here.}:
\bea
\left(\SC^4_{3,3}[\mbox{{\sf WA}$_{n-1}$}]\right)^2 & = & 64\,
\frac{n-3}{n-2}\,\frac{c+2}{5c+22}
\,\frac{c(n+3)+2(4n+3)(n-1)}{c(n+2)+(3n+2)(n-1)} \\[1mm]
{ \SC^4_{4,4}[\mbox{{\sf WA}$_{n-1}$}] \,\,\SC^4_{3,3}[\mbox{{\sf
WA}$_{n-1}$}]}
& = & \nonumber\\[1mm]
&&\hspace{-3cm}{\frac{48}{n-2}\,\frac{c^2(n^2-19) + 3c(6n^3 - 25n^2
+ 15) + 2(n-1)(6n^2-41n-41)}{(5c+22)\ (c(n+2) + (3n+2)(n-1))}}
\ena
for $n = 5$
in agreement with our first solution.

For the second  solution (see Table II)
we find that a  difficulty arises: In  the course of
the
computation it turns out  that in this solution a ladder of null-fields  occur
in the embedding  algebra for any value of the  central charge, starting  at
dimension 8 and therefore already occuring in the
ansatz~(\ref{ans33})-(\ref{ans55}). At this dimension  we find two null-fields,
\bea
N^1_8 & = & P^{3,4}_8 \\[1mm]
N^2_8 & = & P^{3,3}_8 + \kappa_1\, P^{3,5}_8 + \kappa_2 \,P^{4,4}_8
\ena
with
\bea
\kappa_1 & = & {\sm{9}{8}\,\frac{\left( -1 + 2\,c \right) \,
\left( 114 + 7\,c \right) }
    {\left( 10 + c \right) \,\left( -870 + 533\,c + 29\,{c^2} \right) }
\,\SC^4_{3,3}\,\SC^5_{3,4}}\\[1mm]
\kappa_2 & = & {8 \, \frac{\left( 4 - 5\,c \right) \,\left( 2 + c \right) \,
      \left( 114 + 7\,c \right) }
    {\left( 22 + 5\,c \right) \,\left( -870 + 533\,c + 29\,{c^2} \right) }}
\label{kappa}
\ena
We have to take this into account by setting in this solution
$\SC^{P^{3,4}_8}_{n,m}$ and $\SC^{P^{3,3}_8}_{n,m}$ to zero.

The existence of null-fields at any value  of the central charge for the second
solution leads to a remarkable consequence that is in  contrast to  the  first
solution  ({\sf  WA}$_4$):  Even  though  the  null-fields  $N^1_8$  and
$N^2_8$  are taken  from the ansatz~(\ref{ans33})-(\ref{ans55}),    they  still
appear  in some of  the  Jacobi-identities.  Not only  these
lowest-dimensional
null-fields of the algebra violate the Jacobi-identities, but also
null-fields of dimension $> 8$.

This makes  the second  solution similar  to the  exceptional  \W-algebras.
However,  the  differences are obvious:  In exceptional \W-algebras
Jacobi-identity violating null-fields are only null  for a certain value of the
central charge, hence restricting the existence of the algebra to this
$c$-value. In our case the two null-fields $N^1_8$ and $N^2_8$ do not specify a
central charge.

Exceptional \W-algebras can be seen as some ``limit'' of generic  \W-algebras
(f.e.\ the exceptional \W-algebra of the introduction with $T$ and a primary
field of dimension 5/2 can be regarded as  the ``limit'' $c \rightarrow -13/14$
of the super-algebra {\sf WB}(0,2); see also~\cite{HP92}).
It is  not clear whether a
similar point  of view can be taken here (since  in this case there is no
obvious meaning in ``limit'').

Moreover we see that the second solution has the property that the
structure-constants $\SC_{m,n}^k$ behave in the large-$c$ limit as
\be
\left( \SC_{m,n}^k \right)^2 \sim c
\en
that shows that the chosen normalisation ($< W_n(z) W_n(w)  > \,=
{c/n}/{(z-w)^{2n}}$) is not suitable for taking the classical limit. Instead
for the second solution one should normalize these fields such that
\be
\left( \SC_{m,n}^k \right)^2 \sim \mbox{const.} \hspace{1.5cm} \mbox{for $c
\rightarrow \infty$}
\en
showing that this algebra has in the classiqual limit a central term for
the Poisson bracket
$\left\{ T(z), T(w) \right\}$ but not for $\left\{ W_n(z), W_n(w) \right\}$.

\sect{Taking the next step: The {\sf WA}$_5$-algebra}
Finding two solutions for the algebra \W(2,3,4,5) leads automatically to the
question, whether there exist multiple solutions for \W(2,3,4,\ldots,n). We
have
done the computation for \W(2,3,4,5,6) but with a rather disappointing
result: We find
that only one solution exists, namely {\sf WA}$_5$.

In this algebra we also have the choice of a basis for the simple field $W_6$;
we could change the basis by redefining
\be
W_6 \longrightarrow \mbox{norm$(\beta)$} \left( W_6 \,+ \,\beta
P_6^{3,3}\right)
\en
For our calculations we chose a basis in which
\be
<  W_6(z) P_6^{3,3}(w)  > \,\, = \,0
\en
For completeness we give the structure constants in Table 3, with the fields
\be
P_9^{3,3,3} \equiv P_9^{3,P_6^{3,3}} \hspace{1cm} \mbox{and}\hspace{1cm}
P_{10}^{3,3,4} \equiv P_{10}^{3,P_7^{3,4}}
\en

\sect{Null-fields in the algebra \W(2,4,6)}
H.~Kausch and G.~Watts have shown~\cite{KW91} that the  \W-algebra with $T$
and primary fields of dimension 4 and 6,  \W$(2,4,6)$,
has,  similar to the  case considered  here, four non-equivalent
solutions. Redoing  their
calculation  (that was based on four-point-functions) in our frame-work we find
that the similarities between \W(2,4,6) and \W(2,3,4,5) go further.

In the algebra \W(2,4,6) one composite field at dimension 8,  $P^{4,4}_8$ and
two
at dimension 10, $P^{4,4}_{10}$ and $P^{4,6}_{10}$, appear in the irregular
part
of the OPE of the simple fields:
\bea
W_4 \star W_4 & = & \sm{c}{4}\, I + \widetilde{\SC}^4_{4,4} \, W_4 +
\widetilde{\SC}^6_{4,4} \, W_6 \label{ans461}\\[2mm]
W_4 \star W_6 & = & \widetilde{\SC}^4_{4,6} \, W_4 +
\widetilde{\SC}^6_{4,6} \, W_6 +
                     \widetilde{\SC}^{P^{4,4}_8}_{4,6} \, P^{4,4}_8
\label{ans462} \\[2mm]
W_6 \star W_6 & = & \sm{c}{6} \, I + \widetilde{\SC}^4_{6,6} \, W_4 +
\widetilde{\SC}^6_{6,6} \, W_6 +
                     \widetilde{\SC}^{P^{4,4}_8}_{6,6}  P^{4,4}_8 +
\nonumber \\
              &&\widetilde{\SC}^{P^{4,4}_{10}}_{6,6}  \,P^{4,4}_{10} +
                 \widetilde{\SC}^{P^{4,6}_{10}}_{6,6} \, P^{4,6}_{10}
\label{ans463}
\ena
We use $\widetilde{\SC}$ for the structure constants to distinguish them from
those used for \W(2,3,4,5). We refer to~\cite{KW91} for an incomplete list of
them.

In one of the solutions (set 1 of ref.~\cite{KW91}) emerges
a null-field
as  the combination of the two dimension-10 fields~\cite{Bou89}
\be
\widetilde{N}_{10}  \, = \, P^{4,4}_{10} + \kappa P^{4,6}_{10}
\en
with
\be
\kappa \, = \, {\sm{4}{15}\,\frac{\left( -16 + c \right)
\,\left( 24 + c \right) \,
      \left( 21 + 4\,c \right) \,\left( -7 + 10\,c \right) }
    {\left( 64 - 19\,c \right) \,\left( -1 + c \right) \,
      \left( 232 + 11\,c \right) \,\left( -82 + 47\,c + 10\,{c^2} \right) }\,
      \widetilde{\SC}^4_{4,4} \,\widetilde{\SC}^6_{4,4}
      }
\en
The field $\widetilde{N}_{10}$
is again
null for  any value of  the central  charge and as in the case of the
\W$(2,3,4,5)$-algebra it  violates some of  the
Jacobi-identities,  together with  higher-dimensional null-fields of this
algebra.

This special solution of the \W(2,4,6)-algebra is the one that
can be realized by the
Super-Virasoro algebra, identifying (up to normalisations)
\bea
W_4 & \sim & P^{3/2, 3/2}_4 \\
W_6 & \sim & P^{3/2, 3/2}_6
\ena
the spin-3/2 primary field being the supersymmetry generator
$G$~\cite{Bou89,Kau91}. In this realisation the null-field
$\widetilde{N}_{10}$ actually is
identically zero and all Jacobi-identities are zero.

One might think that due to the freedom of modifying the basic OPEs of the
simple fields  by adding the null-field(s), that is  in the case of the
\W(2,3,4,5)-algebra by
\bea
W_4 \stackrel{ \mbox{\tiny (mod)}}{\star} W_5 & = &
W_4 \stackrel{ \mbox{\tiny (orig)}}{\star} W_5 + \alpha_1 N^1_8 +
\alpha_2 N^2_8\\[2mm]
W_5 \stackrel{\mbox{\tiny (mod)}}{\star} W_5 & = &
W_5 \stackrel{ \mbox{\tiny (orig)}}{\star} W_5 + \beta_1 N^1_8 + \beta_2 N^2_8
\ena
and in the case of the \W(2,4,6)-algebra by
\be
W_6 \stackrel{ \mbox{\tiny (mod)}}{\star} W_6 \, = \,
W_6 \stackrel{ \mbox{\tiny (orig)}}{\star} W_6 + \gamma \widetilde{N}_{10}
\en
(the   ``(orig)''  OPEs  refer  to    eqs.~(\ref{ans45}),    (\ref{ans55})
and~(\ref{ans463})  with     $\SC^{P^{3,3}_8}_{4,5}$,
$\SC^{P^{3,4}_8}_{4,5}$,
$\SC^{P^{3,3}_8}_{5,5}$,   $\SC^{P^{3,4}_8}_{5,5}$  and
$\widetilde{\SC}^{P^{4,4}_{10}}_{6,6} $, respectively, set to zero), one
could  adjust the new parameters such that the  Jacobi-identities become zero
exactly. It turns out, however, that this is not the case.

\sect{Conclusions}
Interesting  questions arising  from our  calculations are still  unsolved:
\newline
i) to
find a  realisation for the  second solution of
the algebra \W(2,3,4,5), where the
two null-fields $N^1_8$ and  $N^2_8$ are identically zero (as it  happens in
the
realisations for the second \W(2,4,6)-algebra in  terms of the
Super-Virasoro-algebra);
\newline
ii) even if we  have seen that  \W(2,3,4,5,6) only admits one
solution, {\sf WA}$_5$,
it might be that the two possible solutions to
\W(2,3,4,5) are not a single  accident and including even higher spins might
again
permit more solutions. That leads to the
question, given simple fields of dimensions $2, 3, 4, \ldots, n$, how  many
solutions exist for  a given $n$. We have gained the impression that especially
for odd $n$
several solutions might be possible. If multiple
solutions exist are then all
solutions apart from {\sf WA}$_{n-1}$ ``exceptional'' in the sense that
null-fields appear (in the Jacobi-identities);
\newline
iii) if there  are null-fields in the envelopping algebra of a general
\W-algebra, do these null-fields always violate the Jacobi-identities (either
of the simple fields if their dimension is low enough or the Jacobi-identities
of composite fields)?
\vs{25}

\noindent
Acknowledgement

\noindent
I would like to thank M.~Freeman and E.~Ragoucy for
discussions and K.~Thielemans for
making available his {\it  Mathematica}$^{\mbox{\tiny  TM}}$
package for calculating operator product expansions
and many suggestions how to improve my programs.
Further I would like to
thank H. Kausch for his explanations to \W(2,4,6).

I am grateful to the University of Torino for kind hospitality at the final
steps of this work.
\newpage       

\newpage
\noindent

\begin{center}
\begin{tabular}{|l|c|} \hline\vs{2}
$(\SC^4_{3,3})^2$ & ${\dps\sm{1024}{3}\,\frac{\left( 2 + c \right)
                     \,\left( 23 + c \right) \rule{0 mm}{8 mm}}
                      {\left( 22 + 5\,c \right) \,\left( 68 + 7\,c \right) }}$
                      \\[4mm]
$(\SC^5_{3,4})^2$ & ${\dps 25\,\frac{\left( 116 + 3\,c \right)
                     \,\left( 22 + 5\,c \right) }{\left( 68 + 7\,c \right)
                      \,\left( 114 + 7\,c \right) }}$\\[4mm]
$\SC^{P^{3,3}_6}_{3,5}$ & ${\dps\sm{9}{40}\,\frac{\left( -1 + 2\,c \right)
                           \,\left( 68 + 7\,c \right) }
                            {\left( 2 + c \right) \,\left( 23 + c \right) \,
      \left( 116 + 3\,c \right) }\,\SC^4_{3,3}\,
      \SC^5_{3,4}}$\\[4mm]
$\SC^4_{4,4}$ & ${\dps\sm{9}{32}\,\frac{\left( -128 + 70\,c + {c^2}
                 \right) }{\left( 2 + c \right)
                  \,\left( 23 + c \right) }\,\SC^4_{3,3}}$\\[4mm]
$\SC^{P^{3,3}_6}_{4,4}$ & ${\dps\sm{9}{2}\,\frac{\left( 22 + 5\,c \right) }
    {\left( 2 + c \right) \,\left( 23 + c \right) }}$\\[4mm]
$\SC^5_{4,5}$ & ${\dps-\sm{15}{64}\,
                 \frac{\left( 70272 + 9340\,c + 204\,{c^2} + 11\,{c^3}
\right) }
                  {\left( 2 + c \right) \,\left( 23 + c \right) \,
      \left( 114 + 7\,c \right) }\,\SC^4_{3,3} }$\\[4mm]
$\SC^{P^{3,4}_7}_{4,5}$ & ${\dps\sm{3}{80}\,\frac{\left( 68 + 7\,c \right)
                           \,\left( 334 + 37\,c \right) }
                            {\left( 2 + c \right) \,\left( 23 + c \right) \,
      \left( 116 + 3\,c \right) }\,\SC^4_{3,3}\SC^5_{3,4}}$\\[4mm]
$\SC^{P^{3,4}_8}_{4,5}$ & ${\dps\sm{9}{80}\,\frac{\left( 68 + 7\,c \right) }
                           {\left( 2 + c \right)
                            \,\left( 23 + c \right) }\,\SC^4_{3,3} \,
                            \SC^5_{3,4}}$\\[4mm]
$\SC^{P^{3,3}_6}_{5,5}$ & ${\dps\hspace{7mm}\sm{3}{2}\,
      \frac{\left( 1507824 + 248948\,c + 14880\,{c^2} + 181\,{c^3} \right) }
                            {\left( 2 + c \right) \,\left( 23 + c \right) \,
      \left( 116 + 3\,c \right) \,
\left( 114 + 7\,c \right) }\hspace{7mm}}$\\[4mm]
$\SC^{P^{3,3}_8}_{5,5}$ & ${\dps 4\,\frac{\left( -13656 - 306\,c + 11\,{c^2}
\right) }
                   {\left( 2 + c \right) \,\left( 116 + 3\,c \right) \,
      \left( 114 + 7\,c \right) }}$\\[4mm]
$\SC^{P^{3,5}_8}_{5,5}$ & ${\dps-\sm{3}{8}\,\frac{\left( 68 + 7\,c
\right) }{\left( 2 + c \right)
                            \,\left( 23 + c \right) }\,\SC^4_{3,3}\,
\SC^5_{3,4}}$\\[4mm]
$\SC^{P^{4,4}_8}_{5,5}$ & ${\dps 64\,\frac{\left( 114 + 7\,c \right) }
    {\left( 116 + 3\,c \right) \,\left( 22 + 5\,c \right)\vs{3} }}$\\ \hline
\end{tabular}
\end{center}

\centerline{Table I: {\sf WA}$_4$-solution to \W(2,3,4,5)}
\newpage

\noindent

\begin{center}
\begin{tabular}{|l|c|} \hline\vs{2}
$(\SC^4_{3,3})^2$ & ${\dps\sm{16}{3}\,\frac{\left( 2 + c \right)
                     \,{{\left( 10 + c \right) }^2}\,
      \left( -4 + 5\,c \right)\rule{0 mm}{8 mm} }{\left( 7 + c \right)
                     \,\left( -1 + 2\,c \right) \,
      \left( 22 + 5\,c \right) }}$\\[4mm]
$(\SC^5_{3,4})^2$ & ${\dps 25\,\frac{\left( -1 + c \right)
                     \,\left( 13 + c \right) \,
      \left( 22 + 5\,c \right) }
    {\left( -1 + 2\,c \right) \,\left( 114 + 7\,c \right) }}$\\[4mm]
$\SC^{P^{3,3}_6}_{3,5}$ & ${\dps\sm{9}{32}\,\frac{\left( 7 + c \right)
                           \,\left( -1 + 2\,c \right) \,
      \left( 68 + 7\,c \right)}
    {\left( -1 + c \right) \,\left( 2 + c \right) \,
      \left( 13 + c \right) \,\left( -4 + 5\,c \right) } \, \SC^4_{3,3}
                           \, \SC^5_{3,4}}$\\[4mm]
$\SC^4_{4,4}$ & ${\dps\sm{9}{4}\,\frac{\left( -64 - 6\,c + 45\,{c^2} +
5\,{c^3} \right)}
                 {\left( 2 + c \right) \,\left( 10 + c \right) \,
      \left( -4 + 5\,c \right) }\, \SC^4_{3,3}}$\\[4mm]
$\SC^{P^{3,3}_6}_{4,4}$ & ${\dps\sm{9}{2}\,\frac{\left( 4 + c \right)
                     \,\left( 22 + 5\,c \right) }
    {\left( 2 + c \right) \,\left( -4 + 5\,c \right) }}$\\[4mm]
$\SC^5_{4,5}$ & ${\dps\sm{15}{8}\,\frac{\left( 7 + c \right)
\,\left( -2304 - 188\,c + 1076\,{c^2} +
        85\,{c^3} \right) }
    {\left( 2 + c \right) \,\left( 10 + c \right) \,
      \left( -4 + 5\,c \right) \,\left( 114 + 7\,c \right) }
\,\SC^4_{3,3}}$\\[4mm]
$\SC^{P^{3,4}_7}_{4,5}$ & ${\dps\sm{3}{10}\,\frac{\left( 7 + c \right)
\,\left( -1 + 2\,c \right) \,
      \left( 26 + 7\,c \right) }
    {\left( -1 + c \right) \,\left( 2 + c \right) \,
      \left( 10 + c \right) \,\left( -4 + 5\,c \right) }
\,\SC^4_{3,3}\, \SC^5_{3,4}}$\\[4mm]
$\SC^{P^{3,4}_8}_{4,5}$ & 0\\[4mm]
$\SC^{P^{3,3}_6}_{5,5}$ & ${\dps\sm{3}{2}
\frac{\left( 7 + c \right) \,\left( -154512 - 26404\,c + 46628\,{c^2} +
        6979\,{c^3} + 259\,{c^4} \right) }
    {\left( -1 + c \right) \,\left( 2 + c \right) \,
      \left( 13 + c \right) \,\left( -4 + 5\,c \right) \,
      \left( 114 + 7\,c \right) }}$\\[4mm]
$\SC^{P^{3,3}_8}_{5,5}$ & 0\\[4mm]
$\SC^{P^{3,5}_8}_{5,5}$ & ${\dps\hspace{7mm}\sm{3}{4}
\frac{\left( 7 + c \right) \,\left( -1 + 2\,c \right) \,
      \left( -145776 - 32276\,c + 3222\,{c^2} + 1146\,{c^3} + 49\,{c^4}
\right
        ) }{\left( -1 + c \right) \,\left( 2 + c \right) \,
      \left( 10 + c \right) \,\left( 13 + c \right) \,
      \left( -4 + 5\,c \right) \,\left( -870 + 533\,c + 29\,{c^2} \right) }
\,\SC^4_{3,3}\, \SC^5_{3,4}\hspace{7mm}}$\\[4mm]
$\SC^{P^{4,4}_8}_{5,5}$ & ${\dps 8 \frac{\left( 7 + c \right)
\,\left( 2568 + 2412\,c + 914\,{c^2} +
        35\,{c^3} \right) }{\left( -1 + c \right) \,\left( 22 + 5\,c
\right) \,
      \left( -870 + 533\,c + 29\,{c^2} \right) \vs{3}}}$\\ \hline
\end{tabular}
\end{center}

\centerline{Table II: Solution to \W(2,3,4,5) involving null-fields}
\newpage
\noindent

\begin{center}
\begin{tabular}{|l|c|} \hline\vs{2}
$(\SC_{3,3}^4)^2$       &   $  {{108\,\left( 2 + c \right) \,\left( 30 + c
\right) }\over
    {\left( 25 + 2\,c \right) \,\left( 22 + 5\,c \right) }}$\\[4mm]
$(\SC_{3,4}^5)^2$       &   $ {{125\,\left( 51 + c \right) \,\left(
22 + 5\,c
\right) }\over
    {4\,\left( 25 + 2\,c \right) \,\left( 114 + 7\,c \right) }}$ \\[4mm]
$(\SC_{3,5}^6)^2$       &   $ {{3\,\left( 19 + c \right) \,\left( 11 + 2\,c
\right) \,
      \left( -5 + 4\,c \right) \,\left( 114 + 7\,c \right) \,
      \left( 820 + 11\,c \right) }\over
    {4\,\left( 25 + 2\,c \right) \,
      \left( -329220 + 205197\,c + 43943\,{c^2} + 2052\,{c^3} + 28\,{c^4}
\right) }}$ \\[4mm]
$\SC_{3,5}^{P_6^{3,3}}$ &   $ {{3\,\left( -1 + 2\,c \right)
\,\left( 25 + 2\,c
\right) \,
      \left( -8 + 7\,c \right) \,\left( 68 + 7\,c \right) }\over
    {20\,\left( 2 + c \right) \,
      \left( -329220 + 205197\,c + 43943\,{c^2} + 2052\,{c^3} + 28\,{c^4}
\right) }}\,\,\SC_{3,3}^4\,\SC_{3,4}^5$ \\[4mm]
$\SC_{3,6}^{P_7^{3,4}}$ &   $ {{8\,\left( 23 + c \right) \,\left( 25 + 2\,c
\right) \,
      \left( -4 + 5\,c \right) \,\left( -60 + 29\,c \right) }\over
    {5\,\left( 19 + c \right) \,\left( 51 + c \right) \,
      \left( -5 + 4\,c \right) \,\left( 22 + 5\,c \right) \,
      \left( 820 + 11\,c \right) }}\,\,\SC_{3,4}^5\,\SC_{3,5}^6$ \\[4mm]
$\SC_{3,6}^{P_8^{3,4}}$ &   $ {{4\,\left( -1 + 2\,c \right)
\,\left( 25 + 2\,c
\right) \,
      \left( -4 + 5\,c \right) }\over
    {15\,\left( 19 + c \right) \,\left( 11 + 2\,c \right) \,
      \left( -5 + 4\,c \right) \,\left( 22 + 5\,c \right) }}\,\,
\SC_{3,4}^5\,\SC_{3,5}^6$ \\[4mm]
$\SC_{4,4}^4$           &   $ {{-710 + 1233\,c + 17\,{c^2}}\over
    {36\,\left( 2 + c \right) \,\left( 30 + c \right) }}
\,\,\SC_{3,3}^4$ \\[4mm]
$\SC_{4,4}^6$           &   $ {{\left( 25 + 2\,c \right)
\,\left( 22 + 5\,c
\right) }\over
    {135\,\left( 2 + c \right) \,\left( 30 + c \right) }}
\,\,\SC_{3,3}^4\,\SC_{3,4}^5\,\SC_{3,5}^6$ \\[4mm]
$\SC_{4,4}^{P_6^{3,3}}$ &   $ {{15\,\left( 22 + 5\,c \right) \,
      \left( -17084 + 11179\,c + 1967\,{c^2} + 38\,{c^3} \right) }\over
    {4\,\left( 2 + c \right) \,\left( -329220 + 205197\,c + 43943\,{c^2} +
        2052\,{c^3} + 28\,{c^4} \right) }}$ \\[4mm]
$\SC_{4,5}^5$            &   $ {{5\,\left( -197580 + 2104\,c + 3009\,{c^2} +
11\,{c^3} \right) }\over
    {72\,\left( 2 + c \right) \,\left( 30 + c \right) \,
      \left( 114 + 7\,c \right) }}\,\,\SC_{3,3}^4$ \\[4mm]
$\SC_{4,5}^{P_7^{3,4}}$ &   $ {{\left( 25 + 2\,c \right) \,\left( 258 + 31\,c
\right) }\over
    {15\,\left( 2 + c \right) \,\left( 30 + c \right) \,
      \left( 51 + c \right) }}\,\,\SC_{3,3}^4\,\SC_{3,4}^5$ \\[4mm]
$\SC_{4,5}^{P_8^{3,4}}$ &   $ {{16\,\left( 25 + 2\,c \right) }\over
    {45\,\left( 2 + c \right) \,\left( 30 + c \right) }}
\,\,\SC_{3,3}^4\,\SC_{3,4}^5$ \\[4mm]
$\SC_{4,6}^6$           &   $
{{36895198800 - 20434407460\,c - 6513319448\,{c^2}
- 573770191\,{c^3} -
      21699119\,{c^4} - 409786\,{c^5} - 5096\,{c^6}}\over
    {432\,\left( 2 + c \right) \,\left( 30 + c \right) \,
      \left( -329220 + 205197\,c + 43943\,{c^2} + 2052\,{c^3} + 28\,{c^4}
\right) }}\,\,\SC_{3,3}^4$ \\[4mm]
$\SC_{4,6}^{P_6^{3,3}}$ &   $ {{\left( -1 + 2\,c \right) \,\left( 25 + 2\,c
\right) \,
      \left( 68 + 7\,c \right) \,\left( -32 + 17\,c \right) }\over
    {20\,\left( 2 + c \right) \,
      \left( -329220 + 205197\,c + 43943\,{c^2} + 2052\,{c^3} + 28\,{c^4}
\right) }}\,\,\SC_{3,4}^5\,\SC_{3,5}^6$ \\[4mm]
$\SC_{4,6}^{P_8^{3,3}}$ &   $ {{4\,\left( 25 + 2\,c \right) \,
      \left( -12952440 + 15248466\,c - 4962753\,{c^2} - 491614\,{c^3} -
        8517\,{c^4} + 58\,{c^5} \right) }\over
    {15\,\left( 2 + c \right) \,\left( 19 + c \right) \,
      \left( 51 + c \right) \,\left( 11 + 2\,c \right) \,
      \left( -5 + 4\,c \right) \,\left( 114 + 7\,c \right) \,
      \left( 820 + 11\,c \right) }}\,\,\SC_{3,4}^5\,\SC_{3,5}^6$ \\[4mm]
$\SC_{4,6}^{P_8^{3,5}}$ &   $ {{5\,\left( 25 + 2\,c \right)
\,\left( 22 + 5\,c
\right) \,
      \left( -1530360 + 971226\,c + 177569\,{c^2} + 10851\,{c^3} +
        214\,{c^4} \right) }\over
    {6\,\left( 2 + c \right) \,\left( 19 + c \right) \,
      \left( 30 + c \right) \,\left( 11 + 2\,c \right) \,
      \left( -5 + 4\,c \right) \,\left( 114 + 7\,c \right) \,
      \left( 820 + 11\,c \right) }}\,\,\SC_{3,3}^4\,\SC_{3,5}^6$ \\[4mm]
$\SC_{4,6}^{P_8^{4,4}}$ &   $ {{64\,\left( 25 + 2\,c \right)
\,\left( -4 + 5\,c
\right) \,
      \left( -6240 + 6715\,c + 525\,{c^2} + 14\,{c^3} \right) }\over
    {15\,\left( 19 + c \right) \,\left( 51 + c \right) \,
      \left( 11 + 2\,c \right) \,\left( -5 + 4\,c \right) \,
      \left( 22 + 5\,c \right) \,\left( 820 + 11\,c \right) }}
\,\,\SC_{3,4}^5\,\SC_{3,5}^6$ \\[4mm]
$\SC_{4,6}^{P_9^{3,5}}$ &   $ {{5\,\left( 25 + 2\,c \right)
\,\left( 22 + 5\,c
\right) \,
      \left( -5634 + 3549\,c + 763\,{c^2} + 22\,{c^3} \right) }\over
    {18\,\left( 2 + c \right) \,\left( 19 + c \right) \,
      \left( 30 + c \right) \,\left( 11 + 2\,c \right) \,
      \left( -5 + 4\,c \right) \,\left( 114 + 7\,c \right) }}
\,\,\SC_{3,3}^4\,\SC_{3,5}^6$ \\[4mm]
$\SC_{5,5}^6$           &   $ {{-\left( \left( 25 + 2\,c \right) \,
        \left( 31890 + 221\,c + 13\,{c^2} \right)  \right) }\over
    {675\,\left( 2 + c \right) \,\left( 30 + c \right) \,
      \left( 51 + c \right) }}
\,\,\SC_{3,3}^4\,\SC_{3,4}^5\,\SC_{3,5}^6$ \vs{3}\\ \hline
\end{tabular}
\end{center}

\centerline{Table III: The algebra {\sf WA}$_5$ }

\newpage

\begin{center}
\begin{tabular}{|l|c|} \hline\vs{2}
$\SC_{5,5}^{P_6^{3,3}}$ &   $
{{3\,\left( -368955120 + 190082812\,c + 70048428\,
{c^2} + 9193771\,{c^3} +
        478131\,{c^4} + 5278\,{c^5} \right) }\over
    {4\,\left( 2 + c \right) \,\left( 114 + 7\,c \right) \,
      \left( -329220 + 205197\,c + 43943\,{c^2} + 2052\,{c^3} + 28\,{c^4}
\right) }}$ \\[4mm]
$\SC_{5,5}^{P_8^{3,3}}$ &   $
{{5\,\left( -14058 - 213\,c + 13\,{c^2} \right)
}\over
    {3\,\left( 2 + c \right) \,\left( 51 + c \right) \,
      \left( 114 + 7\,c \right) }}$ \\[4mm]
$\SC_{5,5}^{P_8^{3,5}}$ &   $ {{4\,\left( -138 + c \right) \,\left(
25 + 2\,c
\right) }\over
    {15\,\left( 2 + c \right) \,\left( 30 + c \right) \,
      \left( 51 + c \right) }}\,\,\SC_{3,3}^4\,\SC_{3,4}^5$ \\[4mm]
$\SC_{5,5}^{P_8^{4,4}}$ &   $ {{28\,\left( 114 + 7\,c \right) }\over
    {\left( 51 + c \right) \,\left( 22 + 5\,c \right) }}$ \\[4mm]
$\SC_{5,6}^{P_7^{3,4}}$ &   $ {{\left( 25 + 2\,c \right)
\,\left( -1992448800 +
1293539040\,c +
        219488862\,{c^2} + 15679475\,{c^3} +
449481\,{c^4} + 2482\,{c^5}
\right) }\over
    {9\,\left( 2 + c \right) \,\left( 19 + c \right) \,
      \left( 30 + c \right) \,\left( 51 + c \right) \,
      \left( -5 + 4\,c \right) \,\left( 114 + 7\,c \right) \,
      \left( 820 + 11\,c \right) }}
\,\,\SC_{3,3}^4\,\SC_{3,5}^6$ \\[4mm]
$\SC_{5,6}^{P_8^{3,4}}$ &   $ {{5\,\left( 25 + 2\,c \right) \,
      \left( -1943148 + 1149796\,c + 290443\,{c^2} + 26723\,{c^3} +
        542\,{c^4} \right) }\over
    {162\,\left( 2 + c \right) \,\left( 19 + c \right) \,
      \left( 30 + c \right) \,\left( 11 + 2\,c \right) \,
      \left( -5 + 4\,c \right) \,\left( 114 + 7\,c \right) }}
\,\,\SC_{3,3}^4\,\SC_{3,5}^6$ \\[4mm]
$\SC_{5,6}^{P_9^{3,4}}$ &   $ {\scriptstyle \left( 1603235894400
- 808910372880\,c -
        346356824292\,{c^2} - 40060358452\,{c^3} - 1693452087\,{c^4} -
        19972168\,{c^5} + 297585\,{c^6} + 5894\,{c^7} \right) \times }$
\\
&${{\left( 25 + 2\,c \right) }
\over
    {54\,\left( 2 + c \right) \,\left( 19 + c \right) \,
      \left( 24 + c \right) \,\left( 30 + c \right) \,\left( 51 + c
\right) \,
      \left( 11 + 2\,c \right) \,\left( -5 + 4\,c \right) \,
      \left( 114 + 7\,c \right) \,\left( 820 + 11\,c \right) }}
\,\,\SC_{3,3}^4\,\SC_{3,5}^6 \hspace{4cm}$ \\[4mm]
$\SC_{5,6}^{P_9^{3,6}}$ &   $ {{\left( 25 + 2\,c \right) \,\left(
533746800 -
313683084\,c -
        84091896\,{c^2} - 5725445\,{c^3} - 145113\,{c^4} - 1162\,{c^5}
\right)
      }\over {30\,\left( 2 + c \right) \,\left( 30 + c \right) \,
      \left( 51 + c \right) \,\left( -329220 + 205197\,c + 43943\,{c^2} +
        2052\,{c^3} + 28\,{c^4} \right) }}\,\,\SC_{3,3}^4\,\SC_{3,4}^5$
\\[4mm]
$\SC_{5,6}^{P_9^{4,5}}$ &   $ {{16\,\left( 25 + 2\,c \right) \,
      \left( -6788820 + 4018837\,c + 1080178\,{c^2} + 55937\,{c^3} +
        938\,{c^4} \right) }\over
    {15\,\left( 19 + c \right) \,\left( 51 + c \right) \,
      \left( 11 + 2\,c \right) \,\left( -5 + 4\,c \right) \,
      \left( 22 + 5\,c \right) \,\left( 820 + 11\,c \right) }}
\,\,\SC_{3,4}^5\,\SC_{3,5}^6$ \\[4mm]
$\SC_{5,6}^{P_9^{3,3,3,9}}$ &   $  {{45\,\left( 25 + 2\,c \right)
\,\left(
-4 +
5\,c \right) \,
      \left( -900 - 3171\,c + 161\,{c^2} + 10\,{c^3} \right) \,
      \left( -17084 + 11179\,c + 1967\,{c^2} + 38\,{c^3} \right) }\over
    {2\,\left( 2 + c \right) \,\left( 19 + c \right) \,
      \left( 51 + c \right) \,\left( 11 + 2\,c \right) \,
      \left( -5 + 4\,c \right) \,\left( 820 + 11\,c \right) \,
      \left( -329220 + 205197\,c + 43943\,{c^2} + 2052\,{c^3} + 28\,{c^4}
\right) }}\,\,\SC_{3,5}^6$ \\[4mm]
$\SC_{5,6}^{P_{10}^{3,4}}$ &   $ {{
      \left( -61363491840 + 77827893552\,c - 9606380940\,{c^2} -
        8971840268\,{c^3} - 1297226699\,{c^4} - 71219081\,{c^5} -
        1077068\,{c^6} + 29260\,{c^7} + 784\,{c^8} \right) }
\over
    {27\,\left( 2 + c \right) \,\left( 19 + c \right) \,
      \left( 30 + c \right) \,\left( 11 + 2\,c \right) \,
      \left( -5 + 4\,c \right) \,\left( 114 + 7\,c \right) \,
      \left( -329220 + 205197\,c + 43943\,{c^2} + 2052\,{c^3} + 28\,{c^4}
\right) }} {\scriptstyle \times} $ \\
& ${\scriptstyle 5\,\left( 25 + 2\,c \right)}
\,\,\,\SC_{3,3}^4\,\SC_{3,5}^6
\hspace{13.2cm}$\\[4mm]
$\SC_{5,6}^{P_{10}^{3,6}}$ &   $ {{\left( 25 + 2\,c \right)
\,\left( 272183760 -
172836612\,c -
        33330024\,{c^2} - 1407115\,{c^3} - 21927
\,{c^4} - 182\,{c^5} \right) }
     \over {45\,\left( 2 + c \right) \,\left( 30 + c \right) \,
      \left( 51 + c \right) \,\left( -329220 + 205197\,c + 43943\,{c^2} +
        2052\,{c^3} + 28\,{c^4} \right) }}
\,\,\SC_{3,3}^4\,\SC_{3,4}^5$ \\[4mm]
$\SC_{5,6}^{P_{10}^{4,5}}$ &   $ {{16\,\left( 25 + 2\,c \right) \,
      \left( -8553 + 5653\,c + 896\,{c^2} + 28\,{c^3} \right) }\over
    {15\,\left( 19 + c \right) \,\left( 51 + c \right) \,
      \left( 11 + 2\,c \right) \,\left( -5 + 4\,c \right) \,
      \left( 22 + 5\,c \right) }}
\,\,\SC_{3,4}^5\,\SC_{3,5}^6$ \\[4mm]
$\SC_{6,6}^6$           & $\left({\scriptstyle
18580962172608000 -
        25754639206663200\,c + 5811229201788320\,{c^2} +
        2107803427201590\,{c^3} + 162112912861483\,{c^4} + \hspace{1cm}}
\right.$ \\
& $ \hspace{1cm}{\scriptstyle \left.
        6108517472724\,{c^5} + 264846736659\,{c^6} + 12595175148\,{c^7} +
        309340924\,{c^8} + 2836704\,{c^9} +
11648\,{c^{10}} \right)\,\, \times} $  \\
& $ {{\left( 25 + 2\,c \right) \, }\over
    {1215\,\left( 2 + c \right) \,\left( 19 + c \right) \,
      \left( 30 + c \right) \,\left( 51 + c \right) \,
      \left( 11 + 2\,c \right) \,\left( -5 + 4\,c \right) \,
      \left( 820 + 11\,c \right) \,
      \left( -329220 + 205197\,c + 43943\,{c^2} + 2052\,{c^3} + 28\,{c^4}
\right) }} \,\,\SC_{3,3}^4\,\SC_{3,4}^5\,\SC_{3,5}^6 $ \\[4mm]
$\SC_{6,6}^{P_6^{3,3}}$ &   ${\scriptstyle \left( 2936623997760
- 3752367677312\,c +
480427690100\,{c^2} +
        422397487578\,{c^3} + 66463218105\,{c^4} + 4908088773\,{c^5} +
        191159220\,{c^6} + \right.}$\\
& ${\scriptstyle \left. 3626660\,{c^7} + 21616\,{c^8} \right) }\,{{5}\over
    {12\,\left( 2 + c \right) \,
      {{\left( -329220 + 205197\,c + 43943\,{c^2} + 2052\,{c^3} +
           28\,{c^4} \right) }^2}}} $ \vs{3}\\ \hline
\end{tabular}
\end{center}

\centerline{Table III: {\it continued}}
\newpage

\begin{center}
\begin{tabular}{|l|c|} \hline\vs{2}
$\SC_{6,6}^{P_8^{3,3}}$ &   $ {\scriptstyle \left( -11300841937108800 +
14063403584210400\,c -
        1352296279092708\,{c^2} - 1728077252788212\,{c^3} -
        296011339188793\,{c^4} - \right.\hspace{1cm}}$ \\
& ${ \hspace{1cm} \scriptstyle \left. 21186084487121\,{c^5} -
        709798549133\,{c^6} - 8540989055\,{c^7} + 100398106\,{c^8} +
        3404236\,{c^9} + 19880\,{c^{10}} \right)\, \times } $ \\
& $ {{5}\over
    {3\,\left( 2 + c \right) \,\left( 19 + c \right) \,
      \left( 51 + c \right) \,\left( 11 + 2\,c \right) \,
      \left( -5 + 4\,c \right) \,\left( 114 + 7\,c \right) \,
      \left( 820 + 11\,c \right) \,
      \left( -329220 + 205197\,c + 43943\,{c^2} + 2052\,{c^3} + 28\,{c^4}
\right) }\hspace{1.5cm}} $ \\[4mm]
$\SC_{6,6}^{P_8^{3,5}}$ &   $
{ \scriptstyle \left( -992890166976000 +
        1237084794525600\,c - 121398549036960\,{c^2} -
        150739618392734\,{c^3} - 25893911838127\,{c^4} - \right.
\hspace{1.5cm}} $ \\
& $ {\hspace{2cm} \scriptstyle
\left.     1926099854632\,{c^5} - 73052378239\,{c^6} - 1436629792\,{c^7} -
        13963372\,{c^8} - 65744\,{c^9} \right)\,\times } $ \\
& $ { {\left( 25 + 2\,c \right)}
\over
    {30\,\left( 2 + c \right) \,\left( 19 + c \right) \,
      \left( 30 + c \right) \,\left( 51 + c \right) \,
      \left( 11 + 2\,c \right) \,\left( -5 + 4\,c \right) \,
      \left( 820 + 11\,c \right) \,
      \left( -329220 + 205197\,c + 43943\,{c^2} + 2052\,{c^3} + 28\,{c^4}
\right) }}\,\,\SC_{3,3}^4\,\SC_{3,4}^5 \hspace{0.5cm}$ \\[4mm]
$\SC_{6,6}^{P_8^{4,4}}$ &   $ {\scriptstyle
\left( 2800009400796000 - 3430277833977600\,c
+
        266824907278990\,{c^2} + 435922843100763\,{c^3} +
        81243877432669\,{c^4} + \right. \hspace{1.5cm}} $ \\
& $ {\hspace{2cm} \scriptstyle
\left. 6591440693509\,{c^5} + 284635535593\,{c^6} +
        6791963944\,{c^7} + 82361124\,{c^8} + 346528\,{c^9} \right) \times} $
\\
& $ {{2}\over
    {3\,\left( 19 + c \right) \,\left( 51 + c \right) \,
      \left( 11 + 2\,c \right) \,\left( -5 + 4\,c \right) \,
      \left( 22 + 5\,c \right) \,\left( 820 + 11\,c \right) \,
      \left( -329220 + 205197\,c + 43943\,{c^2} + 2052\,{c^3} + 28\,{c^4}
\right) }\hspace{1.5cm}} $ \\[4mm]
$\SC_{6,6}^{P_{10}^{3,3}}$ &   $
{\scriptstyle {\left( 8034038369665656854858387983612938240000000 -
       5845587987803601154130276947855427635200000\,c - \right. }} $ \\
& $ {\scriptstyle {
       18893467175351525528357058519820988021760000\,{c^2} +
       19568910416504319144341575923744286126848000\,{c^3} + }} $ \\
& $ {\scriptstyle {
       2520228586293396348796072403777045553062400\,{c^4} -
       4821317048567401826806187349857992950753600\,{c^5} - }} $ \\
& $ {\scriptstyle {
       1139695215062021412446781644529245574515040\,{c^6} +
       353529227897143294258318980480572248483824\,{c^7} + }} $ \\
& $ {\scriptstyle {
       222176744578824049576933061656975587576144\,{c^8} +
       52604971805827098579642160546858440094412\,{c^9} + }} $ \\
& $ {\scriptstyle {
       7547941989881264360308312799281405011094\,{c^{10}} +
       737016414477363272201642630180676505699\,{c^{11}} + }} $ \\
& $ {\scriptstyle {
       51252752005528762083427577708182057761\,{c^{12}} +
       2573041084930425666898626173718570626\,{c^{13}} + }} $ \\
& $ {\scriptstyle {
       91832438723539711157037564564257258\,{c^{14}} +
       2167939856521534784877436954876615\,{c^{15}} + }} $ \\
& $ {\scriptstyle {
       24169923357723869973741250003557\,{c^{16}} -
       377180583634502280899116259904\,{c^{17}} -
       22947317521687019788768157662\,{c^{18}} - }} $ \\
& $ {\scriptstyle {
       504355207308191411970605720\,{c^{19}} -
       6217588100833030815408392\,{c^{20}} -
       36440274767963562583872\,{c^{21}} + }} $ \\
& $ {\scriptstyle {\left.
109186439856345385184\,{c^{22}} +
       3796118183735011712\,{c^{23}} + 30125022089653120\,{c^{24}} +
       112091756894208\,{c^{25}} + 164540952576\,{c^{26}} \right) }} $ \\
& $
{{ 15 }\over
   {\left( 19 + c \right) \,\left( 11 + 2\,c \right) \,
     \left( -5 + 4\,c \right) \,{{\left( 114 + 7\,c \right) }^2}\,
     \left( 820 + 11\,c \right) \,
     \left( -329220 + 205197\,c + 43943\,{c^2} + 2052\,{c^3} + 28\,{c^4}
\right) \, d}}
$ \vs{3}\\ \hline
\end{tabular}
\end{center}

\centerline{Table III: {\it continued}}
\newpage

\begin{center}
\begin{tabular}{|l|c|} \hline\vs{2}
$\SC_{6,6}^{P_{10}^{3,5}}$ &   $ {\scriptstyle {
     \left( -1910623477092506587046267455127040000000 +
       331239441064584138264354855016051200000\,c + \right. }} $ \\
& $ {\scriptstyle {
       3660292187357162404301067735341654400000\,{c^2} -
       1288781292257456706478000168638602976000\,{c^3} - }} $ \\
& $ {\scriptstyle {
       1147978357003757557206471221540002728000\,{c^4} +
       33843653058255848358748192210562169120\,{c^5} + }} $ \\
& $ {\scriptstyle {
       169648680445165205524070572320801616560\,{c^6} +
       57550812975394798474941075961950579664\,{c^7} + }} $ \\
& $ {\scriptstyle {
       10423677257560188363999787855626959220\,{c^8} +
       1231029207538218838631458579145791970\,{c^9} + }} $ \\
& $ {\scriptstyle {
       102477524223093276629763383721944141\,{c^{10}} +
       6261430771113011762298739022748213\,{c^{11}} + }} $ \\
& $ {\scriptstyle {
       287261454415938738180017221967320\,{c^{12}} +
       10009960140848905334386050165342\,{c^{13}} + }} $ \\
& $ {\scriptstyle {
       265501790746668749308101133653\,{c^{14}} +
       5315624656748673642279081393\,{c^{15}} + }} $ \\
& $ {\scriptstyle {
       78376289723150701371145066\,{c^{16}} +
       800973019723017815394186\,{c^{17}} +
       4691935526334106047368\,{c^{18}} - }} $ \\
& $ {\scriptstyle { \left.
1130452372695103696\,{c^{19}} -
       285973687077593472\,{c^{20}} - 2467933562943392\,{c^{21}} -
       9581715859968\,{c^{22}} - 14465138688\,{c^{23}} \right) }} $ \\
& $
{{2 \, \left( 25 + 2\,c \right) }\over
   {15\,\left( 2 + c \right) \,\left( 19 + c \right) \,
     \left( 30 + c \right) \,\left( 51 + c \right) \,
     \left( 11 + 2\,c \right) \,\left( -5 + 4\,c \right) \,
     \left( 114 + 7\,c \right) \,\left( 820 + 11\,c \right) \,
d }}
 \,\,\SC_{3,3}^4\,\SC_{3,4}^5$ \\[4mm]
$\SC_{6,6}^{P_{10}^{4,4}}$ &   $
{\scriptstyle{\left( -461226210388844440316682123135498240000000 +
       390546186352260467905385735274832435200000\,c + \right. }} $ \\
& $ {\scriptstyle {
       875512442472005153742474294184519157760000\,{c^2} -
       908241745934270059191644425263063376128000\,{c^3} - }} $ \\
& $ {\scriptstyle {
       163476804170660443812774221187634899110400\,{c^4} +
       225536250748804813274344358005827906737600\,{c^5} + }} $ \\
& $ {\scriptstyle {
       65595227354813361914011699173323857619520\,{c^6} -
       14050524811204270318750310227143532883184\,{c^7} - }} $ \\
& $ {\scriptstyle {
       11195053301919585490988902970960671241528\,{c^8} -
       2926016147658516862369783503200464532232\,{c^9} - }} $ \\
& $ {\scriptstyle {
       458588078515815314117355449019379019730\,{c^{10}} -
       49190438118377185304205783171624374582\,{c^{11}} - }} $ \\
& $ {\scriptstyle {
       3816519673901649735174126957891146067\,{c^{12}} -
       220025327779382505664583255505273888\,{c^{13}} - }} $ \\
& $ {\scriptstyle {
       9528267248125608129528868996995608\,{c^{14}} -
       309081986456745687254649534151146\,{c^{15}} - }} $ \\
& $ {\scriptstyle {
       7353371892273528389128192703511\,{c^{16}} -
       120777461887800900687583954832\,{c^{17}} -
       1106537075617955856920760060\,{c^{18}} + }} $ \\
& $ {\scriptstyle {
       2608343466368955898151328\,{c^{19}} +
       258142779926122049628784\,{c^{20}} +
       4215398180665733980032\,{c^{21}} + }} $ \\
& $ {\scriptstyle {\left.
38460207974174434752\,{c^{22}} +
       211741709240122880\,{c^{23}} + 654042046887936\,{c^{24}} +
       860675751936\,{c^{25}} \right)}} $ \\
& $
{{ 5}\over
   {3\,\left( 19 + c \right) \,\left( 11 + 2\,c \right) \,
     \left( -5 + 4\,c \right) \,\left( 22 + 5\,c \right) \,
     \left( 820 + 11\,c \right) \,
     \left( -329220 + 205197\,c + 43943\,{c^2} + 2052\,{c^3} + 28\,{c^4}
\right) \, d }}
$ \vs{3} \\ \hline
\end{tabular}
\end{center}

\centerline{Table III: {\it continued}}
\newpage

\begin{center}
\begin{tabular}{|l|c|} \hline\vs{2}
$\SC_{6,6}^{P_{10}^{4,6}}$ &   $ {\scriptstyle{
     \left( 225583296756434471408212378368000000 -
       42857521937668998377976174992640000\,c - \right. }} $ \\
& $ {\scriptstyle {
       428476442197814131039216729129152000\,{c^2} +
       154191769779913896531627400813257600\,{c^3} + }} $ \\
& $ {\scriptstyle {
       133824511393646296312914863396779680\,{c^4} -
       4683209857017419979997438001318800\,{c^5} - }} $ \\
& $ {\scriptstyle {
       19914166065042332051607283276221744\,{c^6} -
       6677417171922408421252577328366428\,{c^7} - }} $ \\
& $ {\scriptstyle {
       1197452353663138410737985707144234\,{c^8} -
       139767041265133783832531980850322\,{c^9} - }} $ \\
& $ {\scriptstyle {
       11456045715563103047367396409525\,{c^{10}} -
       685480773541162289713234409538\,{c^{11}} - }} $ \\
& $ {\scriptstyle {
       30591079747101996553714175784\,{c^{12}} -
       1029425846275718856126064434\,{c^{13}} -
       26207185689820356479887945\,{c^{14}} - }} $ \\
& $ {\scriptstyle {
       503004543774436776506790\,{c^{15}} -
       7202157218100242255240\,{c^{16}} - 75420476329430867792\,{c^{17}} -
}} $ \\
& $ {\scriptstyle { \left.
       558064936850199120\,{c^{18}} - 2743141364140000\,{c^{19}} -
       7931555444736\,{c^{20}} - 9944782848\,{c^{21}} \right)}} $ \\
& $
{{16\,\left( 25 + 2\,c \right) }\over
   {5\,\left( 19 + c \right) \,\left( 51 + c \right) \,
     \left( 11 + 2\,c \right) \,\left( -5 + 4\,c \right) \,
     \left( 22 + 5\,c \right) \,\left( 820 + 11\,c \right) \,
d }}
 \,\,\SC_{3,4}^5\,\SC_{3,5}^6$ \\[4mm]
$\SC_{6,6}^{P_{10}^{5,5}}$ &   $
{\scriptstyle{\left( -6729995819753525502520457472000000 +
       2354538709570976947649069372160000\,c + \right. }} $ \\
& $ {\scriptstyle {
       12374940122890321944859761438144000\,{c^2} -
       6522795307754095148734901959152000\,{c^3} -  }} $ \\
& $ {\scriptstyle {
       2951318859518901976272429385381920\,{c^4} +
       585813437899245971854775789513472\,{c^5} + }} $ \\
& $ {\scriptstyle {
       498747323198824608694778178155032\,{c^6} +
       123335744533581103842431140526380\,{c^7} +
       17503551210116860975519900587020\,{c^8} + }} $ \\
& $ {\scriptstyle {
       1653350400190859419656974821683\,{c^9} +
       110832656860882971512572109734\,{c^{10}} +
       5451935801059760743299449996\,{c^{11}} + }} $ \\
& $ {\scriptstyle {
       200238498730388936199395296\,{c^{12}} +
       5527733277081308412011291\,{c^{13}} +
       114461323198019157769966\,{c^{14}} + }} $ \\
& $ {\scriptstyle {
       1758975177341395305738\,{c^{15}} + 19645131058285981664\,{c^{16}} +
       153845035701353176\,{c^{17}} + }} $ \\
& $ {\scriptstyle {\left.
794170123740736\,{c^{18}} +
       2401031259648\,{c^{19}} + 3164249088\,{c^{20}} \right)}} $ \\
& $
{{ 150 }\over
   {\left( 19 + c \right) \,\left( -5 + 4\,c \right) \,
     \left( 114 + 7\,c \right) \,\left( 820 + 11\,c \right) \,
d }}
$ \\[4mm]
$\SC_{6,6}^{P_{10}^{3,3,4}}$ &   $ 0 $ \\[5mm]
where& \\[2mm]
$d$ & $ {\scriptstyle {\left( 32264132475165178970764800000 +
       14922659147344533620778816000\,c -
       48446998713561085019584876800\,{c^2} - \right. }} $ \\
& $ {\scriptstyle {
       7185818828253763119712188000\,{c^3} +
       9158179120209264858864287440\,{c^4} +
       4179812878744043916145776152\,{c^5} + }} $ \\
& $ {\scriptstyle {
       836537510122158655574560012\,{c^6} +
       99107329325671396460455030\,{c^7} + 7708501783947049512473573\,{c^8} +
}} $ \\
& $ {\scriptstyle {
       414234376433599838347945\,{c^9} + 15789935884150642911443\,{c^{10}} +
       431373978892687652533\,{c^{11}} + }} $ \\
& $ {\scriptstyle {
8423092342756537880\,{c^{12}} +
       115709518176813704\,{c^{13}} + 1082823894002608\,{c^{14}} + }} $ \\
& $ {\scriptstyle {\left.
       6506451905744\,{c^{15}} + 22311636480\,{c^{16}} + 32288256\,{c^{17}}
\right) }} $ \vs{3}\\ \hline
\end{tabular}
\end{center}

\centerline{Table III: {\it continued}}


\begin{thebibliography}{99}
%
\bibitem{BS92} P.~Bouwknegt and K.~Schoutens, {\it
``\W-Symmetry in Conformal Field
Theory''}, CERN-TH.6583/92, ITP-SB-92-23, to
appear in Physics Reports.
\bibitem{FRS92} L.~Frappat, E.~Ragoucy and P.~Sorba, {\em ``\W-algebras and
superalgebras from constrained WZW models: A group theoretical
classification''}
ENSLAPP-AL-391-92.
\bibitem{FOR91} L.~Feher, L.~O'Raifeartaigh, P.~Ruelle, I.~Tsutsui and A.~Wipf,
{\em ``On the general structure of Hamiltonian reduction of the WZNW theory''},
DIAS-STP-91-29, UDEM-LPN-TH-71-91, ETH-TH-91-53.
\bibitem{BW91} P.~Bowcock and G.M.T.~Watts, \NP{B379} (1992) 63.
\bibitem{Zam86} A.B.~Zamolodchikov, Theor.\ Math.\ Phys.\ {\bf 65} (1986) 347.
\bibitem{Thi91} K.~Thielemans, \IJMP{C2} (1991) 787.
\bibitem{Mat91} S.~Wolfram, ``{\em Mathematica\/}'', Addison-Wesley (1991).
\bibitem{BFK91} R.~Blumenhagen, M.~Flohr, A.~Kliem, W.~Nahm, A.~Recknagel and
                 R.~Varnhagen, \NP{B361} (1991) 255.
\bibitem{Hor92} K.~Hornfeck, \PL{275B} (1992) 355.
\bibitem{FL88} V.A.~Fateev and S.L.~Lykyanov, \IJMP{A3} (1988) 507.
\bibitem{PRS90} C.N.~Pope, L.J.~Romans and X.~Shen, \NP{339B} (1990) 191.
\bibitem{HP92} C.M.~Hull and L.~Palacios, \MPL{A7} (1992) 2619.
\bibitem{KW91} H.G.~Kausch and G.M.T.~Watts, \NP{B354} (1991) 740.
\bibitem{Bou89} P.~Bouwknegt, {\it ``Extended conformal algebras from Kac-Moody
algebras''} in {\it ``Infinite-dimensional Lie algebras and Lie groups''},
ed.~V.~Kac, Proc.~CIRM-Luminy conference 1988, World Scientific Singapore
(1989).
\bibitem{Kau91} H.G.~Kausch, {\em ``Chiral Algebras in Conformal Field
Theory''}, PhD-Thesis at the University of Cambridge (1991).
%
\end{thebibliography}
\end{document}